\documentclass[conference, 12pt]{revtex4-2}
\usepackage{lipsum} 
\usepackage{amsmath} 
\usepackage{braket}\usepackage{graphicx,amssymb,amsmath,amsthm,amsfonts,epsfig,mathtools}

\usepackage[utf8]{inputenc}
\usepackage{graphicx}
\usepackage{dcolumn}
\usepackage{bm}
\usepackage{amsmath}
\usepackage{mathrsfs}
\usepackage{color}
\usepackage{soul}
\usepackage[dvipsnames]{xcolor}
\usepackage{hyperref}
\usepackage{subcaption}

\begin{document}
\title{ Model-Driven Engineering for Quantum Programming: \\A Case Study on Ground State Energy Calculation\\}

\author{Furkan Polat}
\affiliation{Electrical and Electronics Engineering, Koç University, \& Department of Computer Science, University of Antwerp, Antwerp, Belgium\\
fpolat17@ku.edu.tr}
\author{Hasan Tuncer}
\affiliation{Department of Physics and Astronomy, Purdue University, IN, USA\\
htuncer@purdue.edu}
\author{Armin Moin}
\affiliation{Department of Computer Science, University of Colorado, Colorado Springs, CO, USA\\
amoin@uccs.edu}
\author{Moharram Challenger}
\affiliation{Department of Computer Science, University of Antwerp, \& Flanders Make Strategic Research Center, Antwerp, Belgium\\
moharram.challenger@uantwerpen.be}

\begin{abstract}

\newpage

\noindent

\textbf{Abstract:} 

This study introduces a novel framework that brings together two main Quantum Programming methodologies, gate-based Quantum Computing and Quantum Annealing, by applying the Model-Driven Engineering principles. This aims to enhance the adaptability, design, and scalability of quantum programs, facilitating their design and operation across diverse computing platforms. A notable achievement of this research is the development of a mapping method for programs between gate-based quantum computers and quantum annealers, which can lead to the automatic transformation of these programs. Specifically, this method is applied to the Variational Quantum Eigensolver Algorithm and Quantum Annealing Ising Model, targeting ground state solutions. Finding ground-state solutions is crucial for a wide range of scientific applications, ranging from simulating chemistry lab experiments to medical applications, such as vaccine development. The success of this application demonstrates Model-Driven Engineering for Quantum Programming frameworks' practical viability and sets a clear path for Quantum Computing's broader use in solving intricate problems.\\

\end{abstract}

\maketitle

\newpage

\section{Introduction}
\label{sec:introduction}

Various models of computation exist for Quantum Computing (QC), such as quantum circuit models (i.e., Quantum gates) and adiabatic (i.e., quantum annealing), which are well-established ones\cite{Nimbe2021Models}. However, it is already theoretically proven that any quantum circuit algorithm can be transformed into a quantum annealing algorithm with the same time complexity and vice-versa, hence showing they are polynomial equivalent \cite{Aharonov2007Adiabatic}. 

Furthermore, the key technologies for the actual physical implementation of quantum processors comprise superconducting, ultra-cold atoms (e.g., trapped ions), quantum optics, and spin-based. For instance, Linke et al. \cite{Linke+2017} conducted experiments on two 5-qubit quantum computers with the quantum circuit model, which had two different technologies and architectures, namely superconducting and trapped ions. They illustrated that the superconducting system offered faster gate clock speeds and a solid-state platform, whereas the other one featured superior qubits and reconfigurable connections. They concluded that the performance of those systems reflected the topology of connections in the base hardware, thus supporting the idea that quantum software and hardware should ideally be co-designed \cite{Moin+2022a}. 

The Model-Driven Engineering (MDE) paradigm has already been proposed as a reasonable choice to cope with the heterogeneity and complexity observed in the QC domain and its programming \cite{Moin+2022a, AliYue2020, Gemeinhardt+2021}. 

In this paper, the vision of applying MDE techniques for Quantum Programming (QP)\cite{Moin+2022a,Moin-QP-2023} is extended by proposing a new framework and applying it in the context of the ground state energy calculation problem, which has concrete use cases in various domains. This includes material science for developing new materials \cite{Aryasetiawan1997The}, chemistry for creating efficient catalysts \cite{Kresse1993Ab}, pharmaceuticals for advancing drug and vaccine development \cite{AspuruGuzik2012Photonic}, and physics for exploring phenomena like superconductivity and quantum magnetism \cite{Affleck1987Rigorous}. 


This paper is structured as follows: Section~\ref{sec:related_work} provides a review of related work and presents an overview of QC, the MDE paradigm, and the ground-state energy calculation problem. In Section~\ref{sec:proposed_approach}, it is introduced that MDE framework for quantum programs, focusing on the strategy for developing Platform Independent Model for Ground State Calculation. Section~\ref{sec:methodology} outlines our proof of concept work for Model-Driven Engineering for Quantum Programs (MDE4QP). After implementing a ground state solution separately in gate-based QPUs and Quantum Annealers, it demonstrates the results from abstract mapping. Finally, the paper concludes with a summary of our contributions and future research directions in Section~\ref{sec:conclusion}.

\pagebreak

\section{Related Work and Background}
\label{sec:related_work}

Previous research in QC has explored various models of computation and technologies for physical implementation. Aharonov et al. \cite{Aharonov2007Adiabatic} theoretically demonstrated the equivalence between quantum circuit algorithms and quantum annealing algorithms. This foundational work laid the groundwork for understanding the relationship between different QC models.

In terms of physical implementations, Linke et al. \cite{Linke+2017} conducted experiments using two different QC technologies: superconducting qubits and trapped ions. They highlighted the difference in performance and architecture between these systems, emphasizing the importance of co-designing quantum software and hardware.

The MDE paradigm has emerged as a promising approach for managing the complexity and heterogeneity in quantum programming \cite{Moin+2022a, AliYue2020, Gemeinhardt+2021}. Moin et al. \cite{Moin+2022a} proposed using MDE techniques for quantum programming, laying the foundation for our work.

\subsection{Quantum Computing}

QC is a transformative approach to computing that integrates quantum mechanics principles into information processing This field's theoretical foundation was pioneered by Richard Feynman, who proposed the concept of quantum simulation in 1982. Feynman suggested that quantum systems could simulate physical phenomena more efficiently than classical computers, addressing complex quantum mechanical problems \cite{Feynman1982Simulating}.

Significant progress in QC was marked by the development of Shor's algorithm. This algorithm, aimed at prime factorization, runs much faster on quantum computers than the fastest algorithms on classical computers, highlighting the advanced computational capabilities of quantum systems for certain tasks \cite{Shor1994Algorithms}.

Following these early steps, quantum computing (QC) has undergone significant advancements, demonstrating its capability for speed improvements in various applications, a phenomenon known as quantum speedup. For instance, quantum algorithms can search an unsorted database with \(N\) entries in time proportional to \(\sqrt{N}\), as shown by Grover \cite{Grover1997Quantum}, unlike classical computers, which take time proportional to \(N\). Moreover, quantum computers can perform Fourier transforms, invert sparse matrices, and find their eigenvalues and eigenvectors in time that scales polynomially in \(\log_2 N\), as demonstrated by Biamonte et al. \cite{Biamonte2017Quantum}. This contrasts with classical algorithms that scale as \(N \log_2 N\).

The Sycamore processor by Google demonstrated the quantum advantage by performing a specific computational task exponentially faster than the current state-of-the-art classical supercomputers, highlighting the potential of quantum processors to achieve significant speed-ups for certain tasks \cite{Arute2019Quantum}.

The impact of QC spans various domains, from cryptography, where it could break current encryption schemes \cite{Aumasson2017The}, to optimization challenges in logistics and manufacturing \cite{Ajagekar2019Quantum}, and to detailed simulations in chemistry and physics \cite{Motta2021Emerging}. The potential for QC to accelerate drug discovery and materials science highlights its implications across several fields \cite{Cao2018Potential}.

\subsection{Different Approaches in Quantum Computing}

QC has a variety of computational paradigms, each distinguished by unique principles, methodologies, and quantum phenomena exploitation. There are various models of computation for QC, such as quantum circuit/logic gate, adiabatic/annealing, topological, quantum walks, one clean qubit and measurement-based. Since our work focuses on quantum circuit gate model and Annealing model mapping, this subsection provides an overview of the two primary approaches to them.

\subsubsection{Quantum Gate-based Model}

In this model, quantum operations are represented as sequences of quantum gates, each performing a specific transformation on the qubits. These gates include single-qubit gates, which act on individual qubits, and multi-qubit gates, which entangle multiple qubits. By applying sequences of these gates to the qubits in a quantum circuit, complex quantum computations can be performed. Notably, the gate-based model forms the foundation for various quantum algorithms, including Shor's algorithm \cite{Shor1994Algorithms} for prime factorization, Grover's algorithm for database searching \cite{Grover1997Quantum} and Variational Quantum Algorithms (VQA) for optimization \cite{Farhi2014Quantum}, eigenvalue estimation \cite{Peruzzo2014Variational} and machine learning \cite{Preskill2018Quantum}.

\subsubsection{ Quantum Annealing }
 This model originates from the adiabatic theorem to solve optimization problems by evolving an initial quantum state towards the ground state, which represents the problem's optimal solution. This evolution represents the problem's solution without manipulating the qubits through direct operators, marking a difference from the step-by-step operations seen in gate-based computing. AQC's method is inherently suited for various optimization challenges, encoding solutions in the system's lowest energy state \cite{Bergholm2018PennyLane}.
Quantum Annealers efficiently addresses optimization problems such as logistical planning, financial portfolio optimization, and complex scheduling. Quantum annealing can often reach satisfactory solutions faster than classical computational methods by enabling the system to escape local minima and explore more of the solution landscape. This makes it a valuable tool in scenarios where speed is prioritized over absolute precision in finding the lowest possible energy state \cite{Boixo2013Evidence,King2022Coherent}.

\subsection{Challenges in Quantum Computing}

One of the main challenges in quantum computing is maintaining the coherence of quantum states. Qubits are extremely sensitive to their environment. Even small temperature fluctuations, electromagnetic waves, or other interactions with the environment can cause qubits to lose their quantum properties (decoherence) and the information they contain to become corrupted. Quantum noise can also interfere with the operation of qubits, causing errors in calculations \cite{Cory1998EXPERIMENTAL}.

Therefore, robust error correction methods are very important. Quantum error correction is more complex than classical error correction and requires significant additional qubits and computational resources to encode information in such a way that errors can be detected and corrected without disrupting the quantum state \cite{Knill1996Theory}.

Scalability is another important challenge. Building a quantum computer with enough qubits to solve practical problems is a major challenge. Increasing the number of qubits not only requires maintaining consistency across a larger system but also exponentially increases the complexity of error correction. The physical systems used to realize qubits (superconducting circuits, trapped ions, etc.) currently face significant engineering and manufacturing challenges to scale up while maintaining fidelity \cite{Leon2021Materials,Franklin2004Challenges}.

Quantum computers will need to work together with classical computers to perform practical tasks. This integration poses challenges in terms of creating efficient interfaces and communication protocols between quantum and classical systems \cite{Meter2016The}.

Additionally, since quantum platforms take different approaches, their efficiency on different problems may differ. In this case, it will be important to develop a system that can run these platforms together. Our aim is to eliminate heterogeneity through Model-Driven Engineering (MDE).

Quantum computers will need to work together with classical computers to perform practical tasks. This integration poses challenges in terms of creating efficient interfaces and communication protocols between quantum and classical systems \cite{Meter2016The}.

\subsection{Model-Driven Engineering (MDE)}
 MDE is a software development approach that prioritizes creating and using DSMLs to represent the core functionality of a system. DSMLs capture the desired behaviour and properties at a higher level of abstraction instead of focusing on the final code. This focuses on the problem domain, rather than coding specifics, which leads to better communication and understanding among involved developers\cite{Lee2011Implementing}. MDE offers several benefits, including enabling early validation and verification through techniques like model analysis and simulation, ultimately leading to a more robust and efficient software solution. Additionally, MDE can enhance development productivity by automating tasks such as code generation \cite{Idani2020Alliance}.

\subsection{Ground State Calculation}

The Ground State Calculation Problem is fundamental to understanding the physical and chemical properties of systems at the quantum level. Calculating the ground state of quantum systems enables scientists to predict behaviours and interactions, crucial for fields such as material science and drug discovery.

One area where ground-state energy calculations play a pivotal role is in the development of new pharmaceuticals. Quantum chemistry simulations, facilitated by Variational Quantum Algorithms, are essential for discovering Hamiltonian spectra \cite{Peruzzo2014Variational}. These algorithms are key to analyzing reaction rates and understanding catalytic processes, which are vital for creating new drugs. This method allows for the analysis of molecular energies, critical for predicting chemical reactions and interactions, and showcases the potential of near-term quantum hardware to enhance calculations in chemistry and drug development.

\subsubsection{Classical Algorithms for Ground State Calculation}

Classical algorithms for calculating the ground states of the quantum states are essential in understanding how atoms and molecules behave. These methods help scientists find the lowest energy state of quantum systems, which is important for predicting chemical reactions and material properties. Key techniques include the Variational Principle, Hartree-Fock Method \cite{Becke1993A}, Monte Carlo Simulations \cite{Swendsen1987Nonuniversal}. Each traditional method has its own way of approaching the problem, but all aim to provide accurate predictions about the quantum.

\paragraph{Density Matrix Renormalization Group (DMRG)}

The development of the Density Matrix Renormalization Group (DMRG) algorithm marked a significant advancement in computational physics, addressing some of the limitations faced by classical algorithms. DMRG, primarily applied to one-dimensional quantum systems, optimizes the calculation of low-energy eigenstates by iteratively refining the system's state space, focusing on the most relevant degrees of freedom associated with entanglement. This process is significantly more efficient than traditional methods, offering a glimpse into the potential of specialized algorithms in overcoming computational challenges \cite{Schollwoeck2010The}.

\paragraph{Tensor Networks}

Building upon the success of DMRG, tensor networks extend the concept of efficient quantum state representation to higher dimensions. These networks express the quantum state as a product of interconnected tensors, simplifying the representation of complex entanglements and correlations within the system:

\begin{equation}
\begin{aligned}
\Psi_{i_1,i_2,\ldots,i_N} = \sum_{\alpha_1,\alpha_2,\ldots,\alpha_N} A^{[1]}_{i_1;\alpha_1} A^{[2]}_{i_2;\alpha_1\alpha_2} \ldots 
A^{[N-1]}_{i_{N-1};\alpha_{N-2}\alpha_{N-1}} A^{[N]}_{i_N;\alpha_N\alpha_{N-1}}
\end{aligned}
\end{equation}

where $i_k; \alpha_k$ represents indices corresponding to local properties, and the bond indices labeled with $\alpha_k$ can encode correlations such as entanglement. The decomposition is a special case known as the matrix product state (MPS). The $\alpha_k$ indices' maximum number of values, known as the bond size, serves as a quantitative measure of entanglement.

\paragraph{Artificial Neural Networks in Quantum Physics}

The intersection of quantum physics and computational science has further expanded with the application of artificial neural networks (ANNs) to model quantum systems. ANNs, particularly through structures like the Restricted Boltzmann Machine (RBM), have shown promise in capturing the complex multi-particle wave functions of quantum states \cite{Carleo2019, Dash2019}. The RBM employs a layered approach, where connections between visible and hidden units encode the quantum system's properties:

\begin{equation}
\psi_{\text{RBM}} (\mathbf{j}) = \exp\left (\sum_{i=1}^{L} a_i j_i\right) \prod_{i=1}^{M} \cosh\left (b_i + \sum_{k} W_{ik} j_k\right),
\end{equation}

This method,tensor networks, provides a useful tool for simulating quantum systems, Through the adaptation of ANNs, researchers can model systems with a level of efficiency and accuracy that opens new avenues for exploration and understanding in quantum physics \cite{Vicentini2021,Gao2017, Choo2018}.

The progression from classical algorithms through artificial neural networks, reflects the dynamic and interdisciplinary nature of computational research in quantum physics. Notably, NetKet emerges as a pivotal framework in this domain, offering a library for studying many-body quantum systems with machine learning approaches, highlighting the potential of neural network quantum states for quantum state tomography, supervised learning, and ground state searches \cite{Carleo2019}. Further, the explicit demonstration of initial state construction in ANNs using NetKet and validation with IBM Q Experience Platform underscore the practical applicability and accuracy of these models in simulating quantum computations \cite{Dash2019}. With the release of NetKet 3, built on JAX, the toolbox for many-body quantum systems now supports advanced neural network ansätze, GPU/TPU accelerators, and discrete symmetry groups, marking significant advancements in the field \cite{Vicentini2021}.

 Those classical methods encounter challenges as the system's complexity escalates with the addition of more particles. Real-world problems often involve large systems with many interacting particles, and the intricate interactions within such systems result in an exponential growth in computational workload. This presents a significant challenge that classical computers, with their inherent limitations, struggle to manage effectively.
 
Quantum computers, on the other hand, offer an advantage, and Qubits play a key role there. Adding a qubit to the system doubles the computational capacity. So, with n qubits, the computational possibilities expand to $2^n$. This scalability provides quantum computers with an edge, offering a promising solution for handling the complexities of calculating the ground states of the molecules \cite{Abrams1997Simulation,Abrams1999Quantum}.

\paragraph{Quantum Phase Estimation}

Recent advancements in quantum computing have introduced new methods for addressing eigenvalue problems that are challenging for classical computational approaches. A key development in these quantum solutions is the Quantum Phase Estimation (QPE) algorithm, notable for its efficiency advantage over conventional computational methods. The QPE algorithm necessitates \(O (p^{-1})\) quantum operations to achieve a specified level of precision, \(p\), in estimating the eigenvalue linked to a known eigenvector of a Hermitian operator \(H\).

The precision \(p\) highlights the accuracy in the eigenvalue estimation process, with the algorithm assuming initial access to the eigenvector in question. The aim is to uncover the associated eigenvalue. The quantum operations required for this process, directly related to \(O (p^{-1})\), mandate sequential execution of the operation \(e^{-iHt}\). In practical applications, this could require to use of millions or even billions of quantum gates—far exceeding the operational capacity of current quantum technologies, which are predominantly capable of managing only tens to hundreds of gates \cite{Jones2012FasterQuantum,Whitfield2011Simulation,AspuruGuzik2012Photonic}.

\paragraph{Variational Quantum Eigensolver (VQE)}

\begin{figure}
 \centering
 \includegraphics[scale=0.65]{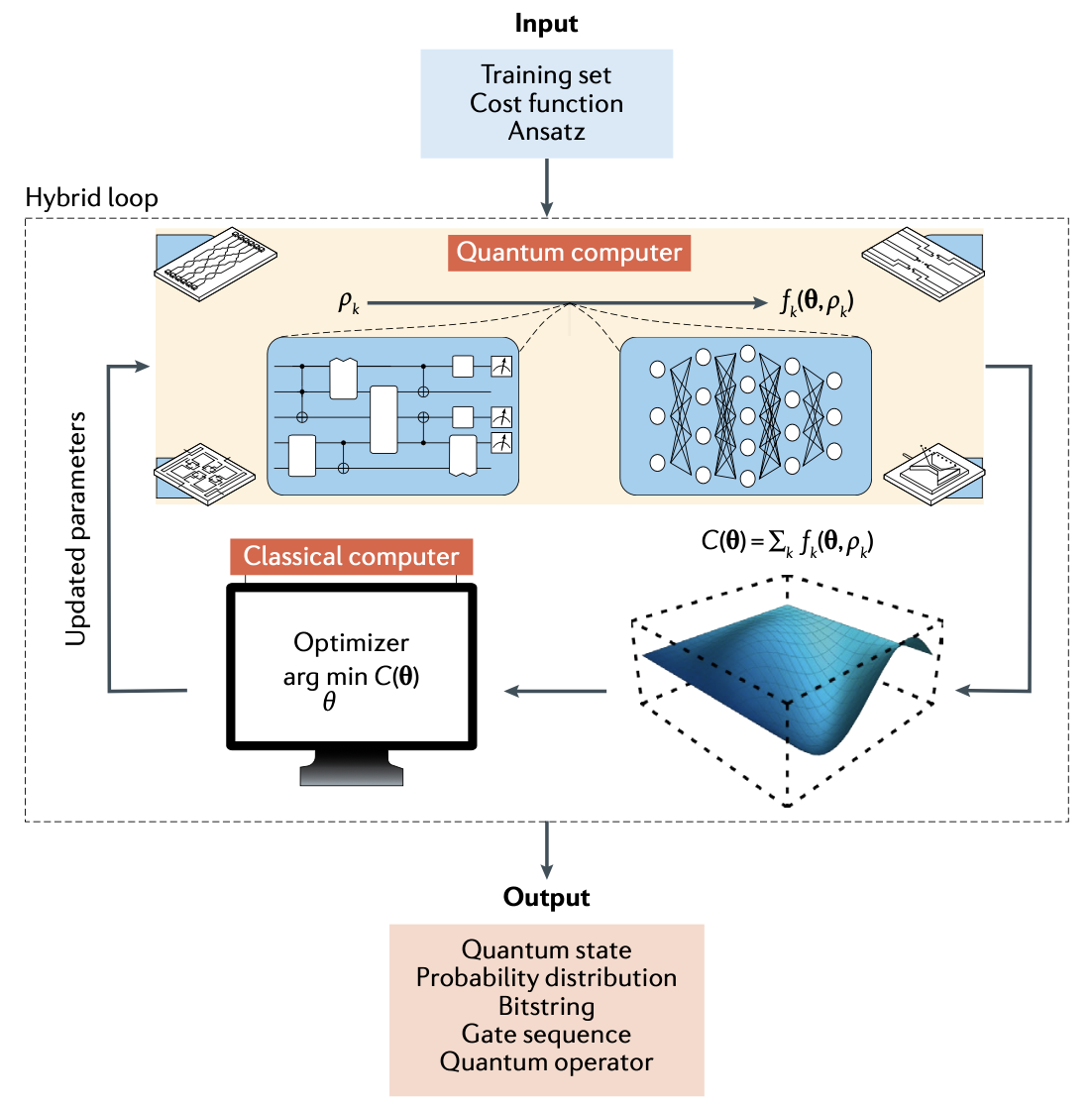}
 \caption{The schematic diagram illustrates a variational quantum algorithm (VQA). \cite{Cerezo2021Variational}}
 \label{Fig:VQA}
\end{figure}

In hybrid VQE solvers, the crucial quantum component involves the utilization of Ising model solvers. Quantum computers work on the Ising model solution based on a given ansatz as illustrated in Fig. \ref{Fig:VQA}. The inputs include a cost function $C (\boldsymbol{\theta})
$, where $\boldsymbol{\theta}$ represents parameters encoding the solution, an ansatz whose parameters are trained to minimize the cost, and potentially a set of training data $\{\rho_k\}$ used in optimization. Typically, the cost function can be expressed as a set of functions $\{f_k\}$. 

The ansatz is depicted as a parameterized quantum circuit on the left, next to a neural network shown on the right. During each iteration, a quantum computer efficiently estimates the cost (or its gradients), with this information fed into a classical computer utilizing optimizers to navigate the cost landscape $\boldsymbol{\theta}$ and solve the optimization problem. Upon meeting termination conditions, the VQE provides an estimate of the solution, with the form of the output dependent on the specific task. The red-shaded box highlights common types of outputs.
is a hybrid algorithm designed to estimate the ground state energy of a molecule by blending variational techniques with quantum mechanics. It operates iteratively, fine-tuning wave function guesses to minimize energy until it reaches the system's most stable state. This process involves both quantum and classical components: the quantum part computes the system's energy for a given Hamiltonian, while the classical part optimizes variational parameters \cite{Peruzzo2014Variational, Cerezo2021Variational}.

The fundamental equation at the heart of VQE for ground sate calculation is given by:
\begin{equation}
 H \ket{\psi} = E \ket{\psi},
\end{equation}
where $H$ represents the Hamiltonian acting on the wave function $\psi$, and $E$ denotes the associated energy.

The Hamiltonian $H = \hat{T} + \hat{V}$ encompasses kinetic and potential energies, incorporating terms that describe electron-electron, nucleus-nucleus, and electron-nucleus interactions for electrons and nuclei. The representation of the wave function evolves from classical Slater determinants to second quantization. This methods expresses the states using fermionic creation and annihilation operators. This formalism facilitates the expression of the Hamiltonian for computation on quantum computers \cite{Weinberg1995The}.

VQE employs techniques like the Jordan-Wigner transformation to map fermionic operators to qubit operators that enables encoding of molecular systems into qubits \cite{Havl'ivcek2017Operator}. The goal is to minimize the expectation value of the Hamiltonian:
\begin{equation}
E_\theta = \min_\theta \frac{\langle \Psi (\theta) | \hat{H} | \Psi (\theta) \rangle}{\langle \Psi (\theta) | \Psi (\theta) \rangle},
\end{equation}

where $\theta$ represents the parameters iteratively optimized through quantum gates. This iterative process aims to closely approximate the ground state of the molecular system, leveraging the precision and efficiency of quantum computation for energy calculations

\paragraph{Quantum Annealing} 
 is also employed for ground state calculations, as it uses the principles of adiabatic quantum mechanics to solve optimization problems. It progresses from an initial Hamiltonian, representing the problem at hand, to a final Hamiltonian that corresponds to the ground state solution.

\paragraph{Ising Model} 

describes interacting spins in a system, where each spin interacts with its neighboring spins on a lattice grid. The interactions are defined by coupling constants, and there may also be local magnetic fields acting on each spin.\cite{Haley1978II} In VQE, the Ising model is used to map the terms of the molecular Hamiltonian onto a set of qubits, enabling quantum computers to simulate molecular systems and calculate their properties efficiently.\cite{Cerezo2021Variational}

Additionally, the Ising model finds widespread application in other contexts, such as in quantum annealers, where optimization problems are initially formulated as Ising models. The ground energy of these Ising models represents the optimal solution for various problems, including the traveling salesman problem and financial optimization.

\pagebreak

\section{Proposed Approach}
\label{sec:proposed_approach}

A well-established MDE could provide a robust framework for developing quantum algorithms because it has shown its ability for automated code generation, model checking, and analysis in other domains. This paper presents a proof of concept for applying MDE to a quantum application to its hybrid/quantum equivalent or vice versa. As a first step, the paper is particularly focusing on ground-state calculations. 

This paper proposes a new method that combines MDE with QC to tackle the complexities of systems with multiple particles. This approach aims to improve upon traditional methods and provide a new perspective for conducting ground-state calculations, where the complexity of systems involving multiple particles often makes traditional computational approaches insufficient\cite{Carleo2016Solving}. With the MDE paradigms that resolve heterogeneity in both classical and QC, this research contributes to advancing QC applications for solving practical problems \cite{moin2021mde4qai,Moin+2022a, AliYue2020, Gemeinhardt+2021}.

The future of Classical and Quantum Computing will likely involve a mix of processors with different architectures and technologies. This will make software development more complex and require more abstraction. Modeling languages and Domain-Specific Modeling Languages (DSMLs) will be essential for providing this abstraction. These methods allow developers to model their applications without concerning about the underlying hardware or software. MDE and DSMLs will also be important for analyzing and verifying these applications, as the platforms will be complex and distributed. This will require formalizing the semantics of QC and the specific target platforms for QC. In other words, the knowledge of QC experts needs to be transferred to the modeling languages and tools.

MDE4QP builds upon the existing MDE tools in classical platforms. Developers can focus on their application's core functionality without getting involved with the intricacies of the underlying hardware or software. MDE4QP aspires to bring this same level of abstraction to QC platforms, particularly applications involving Quantum Machine Learning (QML). 

\begin{figure}
 \centering
 \includegraphics[width=\columnwidth]{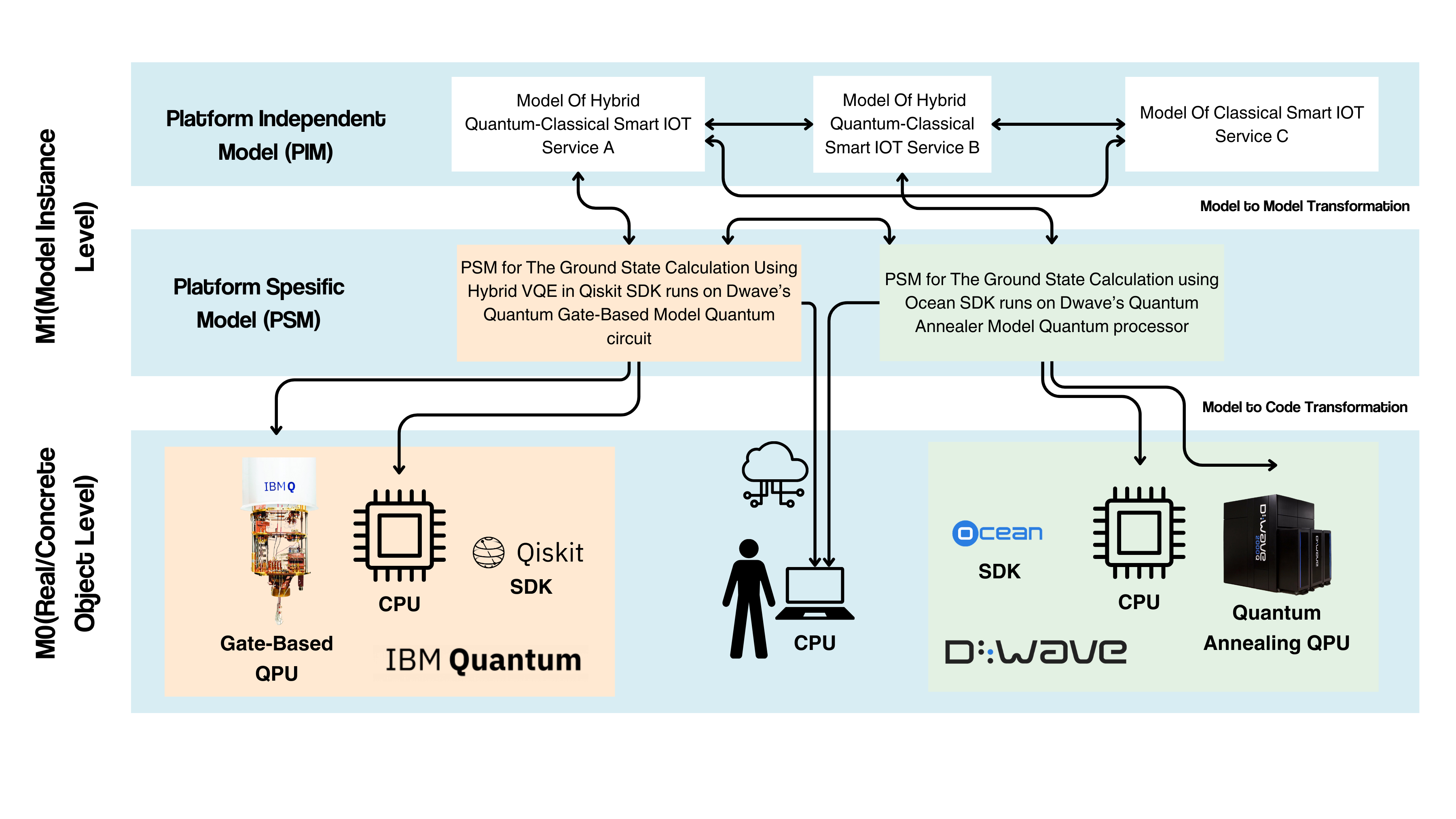}
 \caption{ The Proposed Vision of MDE4QP in the context of Model Driven Architecture for the ground state calculations in different quantum platforms} 
 \label{Fig:Proposed_Vision}
\end{figure}

Similar to how Domain- Specific Modeling (DSM) methodology offers a layer of abstraction, MDE4QP promises to do the same for Quantum program development. This approach is illustrated in Fig. \ref{Fig:Proposed_Vision}. for the proof of concept case in ground state calculation.

The proposed approach prioritizes flexibility and adaptability for hardware platforms. Developers might not know the specific hardware (CPUs, GPUs). Still application will run at runtime once they use DSMLs. MDE4QP aspires to extend this flexibility to the world of QC by covering various Quantum Processor Units (QPUs) with their diverse architectures and capabilities.

 Additionally providing abstraction, MDE4QP aims to streamline the development process further through automated code generation. This will be achieved by implementing model-to-code transformations, essentially translating the high-level models created by developers into code that can run on the target platform. But the potential benefits go beyond mere code generation. The paper explores the intriguing possibility of using model-to-model transformations to achieve various functionalities at the modeling level. These functionalities includes the ability to:

\begin{itemize}
 \item Convert between classical and quantum applications: This feature enables the seamless transition of applications between classical and QC environments, providing flexibility in the choice of computational model.
 
 \item Translate between different models of quantum computation: Such as quantum circuits and quantum annealing. This capability is essential for optimizing applications for specific problem types and uses the strengths of different QC approaches.
 
 \item Fine-tune applications for specific QPU architectures: This feature allows developers to optimize of applications to maximize performance and efficiency on a particular QPU with its unique characteristics. Developers can design applications with tailored optimizations for the target QPU, enhancing overall computational performance.
\end{itemize}

 Also, MDE4QP recognizes the importance of early-stage analysis and verification. The approach aims to enable these activities at the modeling level, allowing developers to catch errors and optimize performance before the actual code is generated. This can significantly reduce development time and effort while ensuring the robustness and efficiency of the final application.

The first step towards the MDE4QP paradigm focuses on establishing a unified abstract mapping to bridge the gap between theoretical quantum algorithms and their practical implementation across various platforms. The work for this framework begins with the conceptual mapping of quantum computations for ground-state calculations. This initial step lays the groundwork for designing and executing algorithms that can seamlessly operate across different platforms.

\begin{figure}
 \centering
 \includegraphics[scale=0.27]{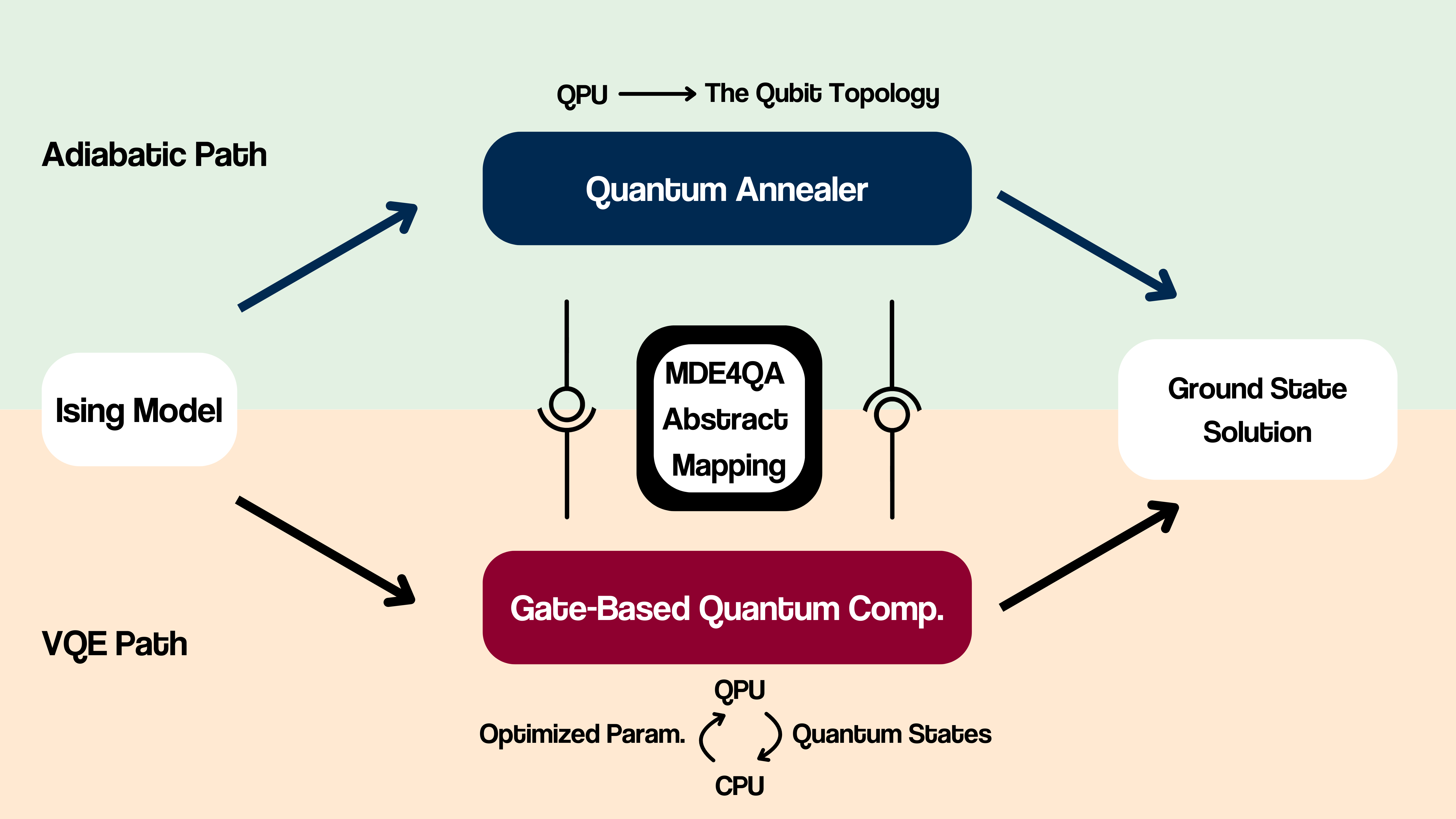}
 \caption{ The schematic diagram illustrates the abstract mapping of solutions from different platforms to ground state calculation. The upper green side represents the Quantum Annealing aproach, while the lower orange side depicts the Hybrid VQE approach in a Quantum Gate-based model. This conceptually simplifies the two methods for solving the same ground state calculation problem." }
 \label{Fig:MDE4AQ}
\end{figure}

There are two paths for calculating he ground states solution of an Ising model: Adiabatic and VQE, as illustrated in \ref{Fig:MDE4AQ}. And Quantum Annealers uses the adiabatic path to evolve the Ising model.On the other side, VQE path includes the gate-based QPU processing of the model. The core objective of the methodology is to implement a Platform Independent Model (PIM) of the Ising model solution, which includes quantum gate-based models and quantum annealer models. By conceptualizing the solutions to the Ising model in a platform-agnostic manner, the flawless transition of quantum algorithms from one computational model to another is enabled

This work explores the relationship between quantum annealing ising model and Quantum Gate-Based VQE through the framework of abstract mapping. This analysis reveals the potential for cross-platform compatibility between different QC paradigms. Such compatibility facilitates the emergence of a new generation of versatile quantum applications. These applications would bridge the gap between diverse hardware platforms and enable the development of adaptive solutions.

\paragraph{ Model-to-model Transformation}

The next step in this paradigm shift involves the design of domain-specific models (DSMs) that facilitate seamless transfer between diverse quantum and classical computing platforms. A system for model-to-model (M2M) transformation across various computation models is envisioned. This system would leverage PIMs (as illustrated in Fig. \ref{Fig:Proposed_Vision}) to translate algorithms for execution on specific quantum hardware. 

The ultimate goal lies in establishing a comprehensive framework that can dynamically adapt to the operational paradigms of both quantum annealing and gate-based architectures. This framework should ensure efficient algorithm translation and execution across platforms, minimizing the need for extensive re-engineering.

\paragraph{Model-to-code Transformation (Automated Code Generation)}

Related to designing Model-to-Model Transformation, Automated Code Generation through Model-to-code transformation is essential. This approach, as shown in Fig. \ref{Fig:Proposed_Vision}, would enable the automatic generation of code from platform-specific models into desired application programming interfaces (APIs).

Automated code generation from high-level models into executable quantum code would significantly accelerate the development of quantum algorithms. This would, in turn, enhance the accessibility of QC for researchers and practitioners who might lack in-depth expertise in quantum programming languages.

To achieve this goal, the development of specialized tools and compilers is necessary. These tools would be responsible for interpreting high-level algorithmic descriptions and generating optimized quantum code tailored for various platforms. These tools will require a level of intelligence to navigate the complexities of quantum computation. This includes tasks like qubit allocation, gate selection, and the incorporation of error correction strategies. Ultimately, the aim is to produce efficient and reliable quantum programs.

\pagebreak

\section{Case Study: Ground State Energy Calculation using Gate-based QC and Quantum Annealer}

\label{sec:methodology}

This research explores a new method for applying the Ising model to both gate-based quantum computers and quantum annealers. This method uses a special mapping process to translate quantum algorithms between these two platforms. This allows researchers to easily switch between different types of quantum computers to solve problems, which can lead to faster discoveries and better solutions.

This method is tested by applying it to three different versions of the Ising model on both gate-based and quantum annealing machines. The goal is to find the lowest possible energy state (ground state) for each version of the model. Here's the Ising model's general Hamiltonian, which describes its energy:

\begin{equation}
 H = \sum_{i,j} J_{ij} \sigma_i^z \sigma_j^z + \sum_{i} h_i \sigma_i^z
 \label{eq::hamiltonyen}
\end{equation}

 where \(J_{ij} = 1\) denotes uniform interaction strength between connected qubits \(i\) and \(j\), \(h_i = 1\) represents a uniform external magnetic field applied to each qubit, and \(\sigma_i^z\) is the Pauli-Z operator acting on qubit \(i\). The choice of \(J_{ij} = 1\) is often made for simplicity and normalization purposes. It allows easier comparison and analysis of different systems. Additionally, this choice simplifies mathematical expressions and calculations while still capturing the essential behavior of the system. 

The specific entanglement patterns are described as follows:

\begin{itemize}
 \item Full-Entangled Ising Model: The model contains interactions among all pairs of qubits, which can be mathematically described as:
 \begin{equation}
 \sum_{i < j} \sigma_i^z \sigma_j^z \quad 
 \end{equation}

 This notation sums over all unique pairs \(i, j\), indicating all-to-all connectivity.

 \item Linear-Entangled Ising Model: This model features a sequential chain of interactions, which can be expressed as:
 
 \begin{equation}
 \sum_{i=0}^{N-2} \sigma_i^z \sigma_{i+1}^z
 \end{equation}
 
 where \(N\) is the total number of qubits, illustrating a linear sequence of connections from qubit \(0\) to \(N-1\).

 \item Circular-Entangled Ising Model: To represent the circular connectivity where the last qubit is also connected back to the first:
 
  \begin{equation}
\sum_{i=0}^{N-1} \sigma_i^z \sigma_{ (i+1) \mod N}^z
 \end{equation}

 Here, \(\mod N\) ensures that after the last qubit, the summation loops back to the first qubit, forming a circular entanglement pattern.
\end{itemize}

For those interested in further exploration or contributions, the project's code is available on GitHub.
\footnote{\url{https://github.com/micss-lab/MDE4QP/}}.

\subsection{VQE with Different Ising Models using Gate-Based Quantum Computing}

 The First step of the investigation highlights the VQE approach to tackle the Ising model with gate-based QC. VQE, a hybrid quantum-classical algorithm, uses the quantum mechanical strengths of superposition and entanglement, combined with classical optimization routines, to efficiently approximate the ground state of quantum systems. 
 
The analysis uses a 1-D Ising model, articulated through Pauli spin matrices, to explore the interplay between quantum spins under the influence of an external magnetic field (\(\lambda\)), set to a constant value of 1 for this study.

This Hamiltonian in \ref{eq::hamiltonyen} serves as the objective function for the VQE algorithm, guiding the search for the system's ground state.

\begin{figure}[!ht]
 \centering
 \begin{subfigure}[b]{0.65\linewidth} 
   \includegraphics[width=\linewidth]{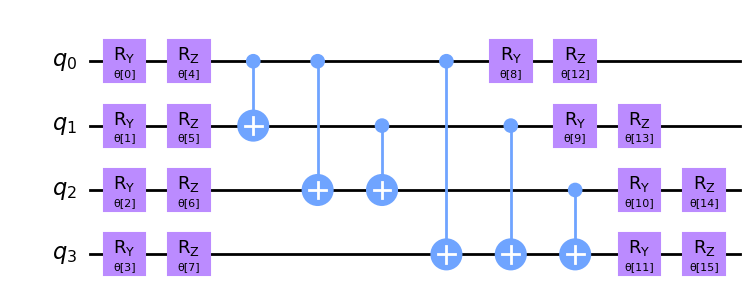}
   \caption{Full-entangled ansatz}
   \label{subfig:full_ent}
 \end{subfigure}
 \hfill 
 \begin{subfigure}[b]{0.65\linewidth}
   \includegraphics[width=\linewidth]{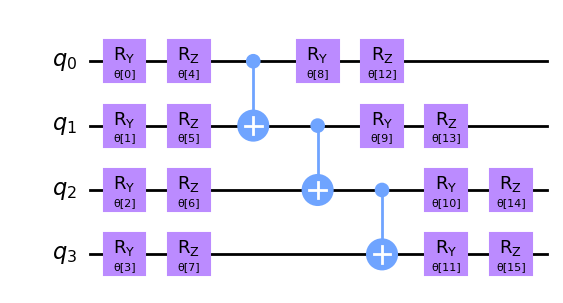}
   \caption{Linear-entangled ansatz}
   \label{subfig:lin_ent}
 \end{subfigure}
 \hfill 
 \begin{subfigure}[b]{0.65\linewidth}
   \includegraphics[width=\linewidth]{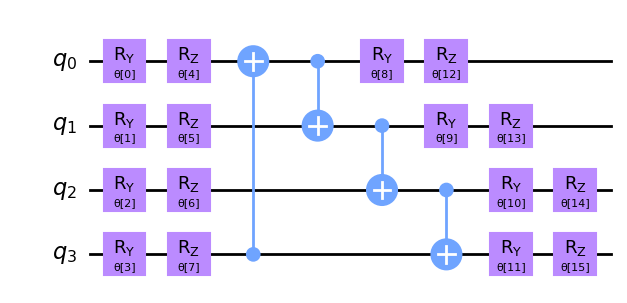}
   \caption{Circular-entangled ansatz}
   \label{subfig:circ_ent}
 \end{subfigure}

 \caption{Circuits of different entangled ansatzes. Each subfigure depicts a unique quantum circuit layout corresponding to its entanglement strategy.}
 \label{fig:ansatz_circuits}
\end{figure}

A 4-qubit system is used to explore various entanglement patterns within the ansatz designs: full-entangled (Fig.~\ref{subfig:full_ent}), linear-entangled (Fig.~\ref{subfig:lin_ent}), and circular-entangled (Fig.~\ref{subfig:circ_ent}). These designs are instrumental in assessing the impact of entanglement on the optimization process and the algorithm's ability to approximate the ground state energy of the corresponding model.These ansatz circuits derived from \cite{Sim2019Expressibility} which were initially used in \cite{Schuld2020Circuit}.
The convergence towards the ground state energy is visually tracked through energy plots for each ansatz, revealing the algorithm's effectiveness in approximating the Ising model's ground state.

\begin{figure}[!ht]
 \centering
 \begin{subfigure}[b]{0.9\linewidth}
   \includegraphics[width=\linewidth]{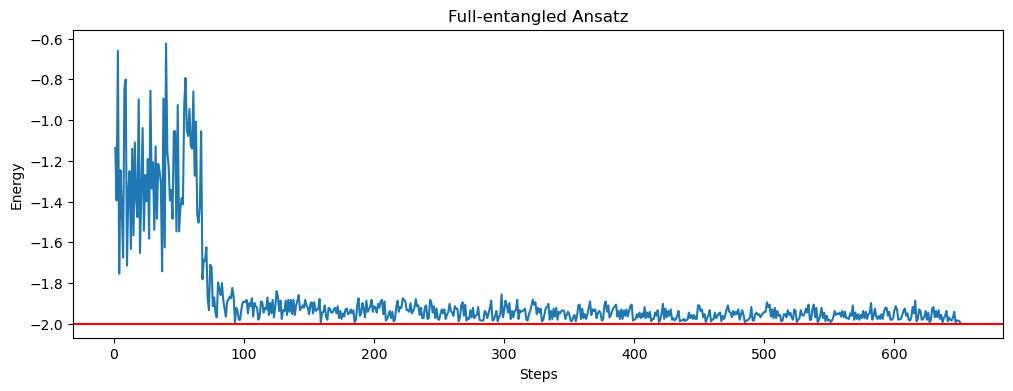}
   \caption{Full-entangled ansatz}
   \label{subfig:full_vqe}
 \end{subfigure}
 
 \begin{subfigure}[b]{0.9\linewidth}
   \includegraphics[width=\linewidth]{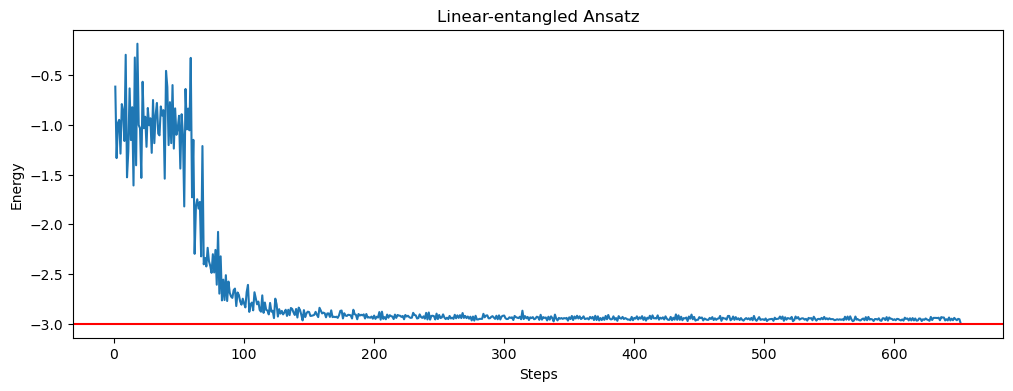}
   \caption{Linear-entangled ansatz}
   \label{subfig:lin_vqe}
 \end{subfigure}
 
 \begin{subfigure}[b]{0.9\linewidth}
   \includegraphics[width=\linewidth]{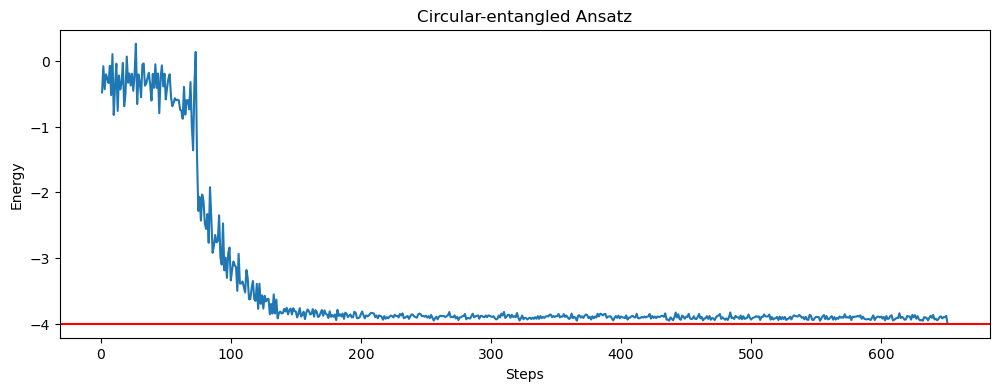}
   \caption{Circular-entangled ansatz}
   \label{subfig:circ_vqe}
 \end{subfigure}

 \caption{Step vs energy plots for different entangled ansatzes. Each subfigure illustrates the energy convergence with the red line indicating the exact ground state energy.}
 \label{fig:vqe_models}
\end{figure}

\begin{itemize}
 \item The \textbf{full-entangled ansatz} reached an energy state of $-2.0$, with the qubit configuration \(\ket{0, 1, 0, 1}\).The iteration steps are shown in figure.~\ref{subfig:full_vqe}.
 \item The \textbf{linear-entangled ansatz} approached the ground state with an energy of $-3.0$, with the qubit configuration \(\ket{0, 1, 0, 1}\).The iteration steps are are shown in figure.~\ref{subfig:lin_vqe}.
 \item The \textbf{circular-entangled ansatz} demonstrated significant promise by converging to a ground state energy value of $-4.0$, with the qubit configuration \(\ket{0, 1, 0, 1}\).The iteration steps shown in figure.~\ref{subfig:cir_vqe}.
 
\end{itemize}

This VQE model investigates the 1-D Transverse Ising model using computational techniques to determine -state energy properties. The investigation via VQE with Qiskit demonstrates convergence to the target ground-state energies through optimization. Across all ansatz types, convergence to the ground-state energies (-2, -3, -4) for corresponding full linear and circular Ising models is observed. 

\subsection{Quantum Annealer Ground State Solution for Different Ising Models}

After gate-based QC, this study also investigates the application of quantum annealing to solve the Ising model using the D-Wave system. Quantum annealing uses quantum mechanical phenomena such as superposition and quantum tunneling, offering a specialized and efficient way to find the system's ground state. This process is inherently suited to solving optimization problems, like finding the minimum energy configuration in the Ising model, by gradually transitioning from a superposition of all possible states towards the lowest energy state.

The D-Wave system facilitates this exploration through its Quantum Processing Unit (QPU), designed around a lattice of qubits that can be programmatically entangled and biased to represent the Ising model. The Ising model's Hamiltonian \ref{eq::hamiltonyen}, representing the energy landscape of the system, is directly mapped onto the QPU. This mapping involves defining the external magnetic field (h) affecting individual qubits and the interaction strength (J) between qubits as circular, linear, and full-entangled models.

\begin{figure}[!ht]
 \centering
 \begin{subfigure}[b]{0.9\linewidth}
   \includegraphics[width=\linewidth]{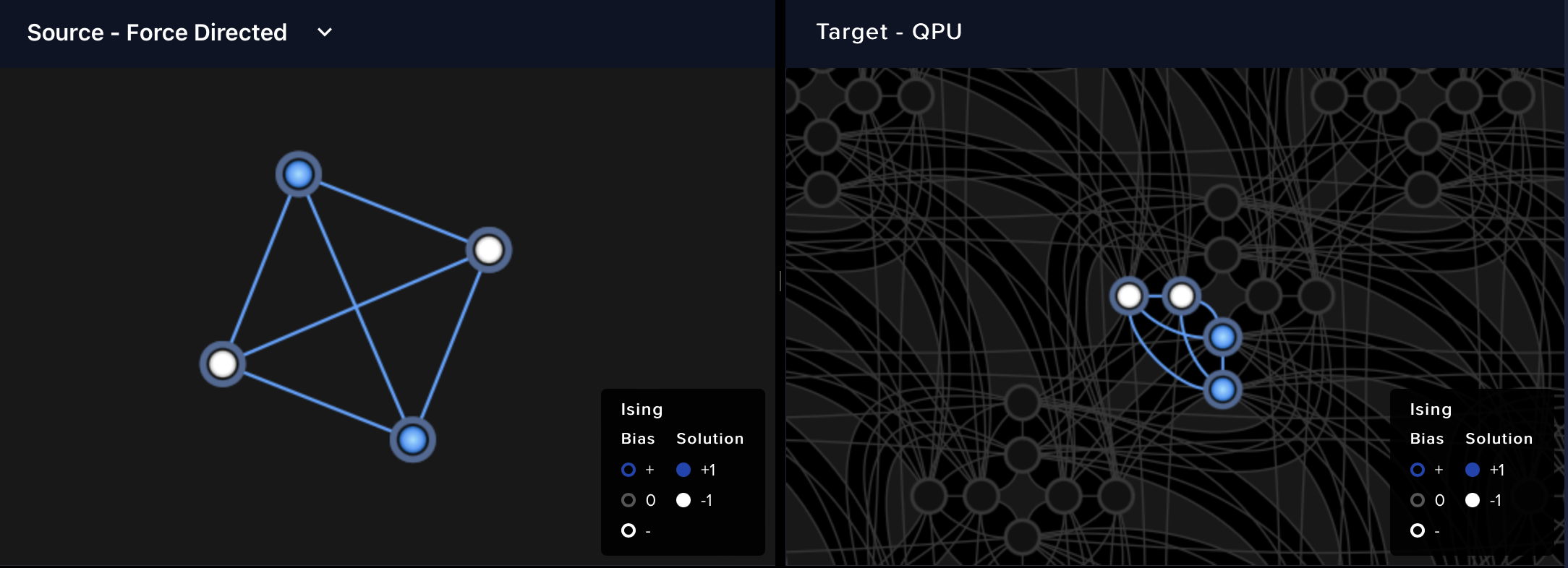}
   \caption{Full-entangled Ising Model}
   \label{subfig:full}
 \end{subfigure}

 \begin{subfigure}[b]{0.9\linewidth}
   \includegraphics[width=\linewidth]{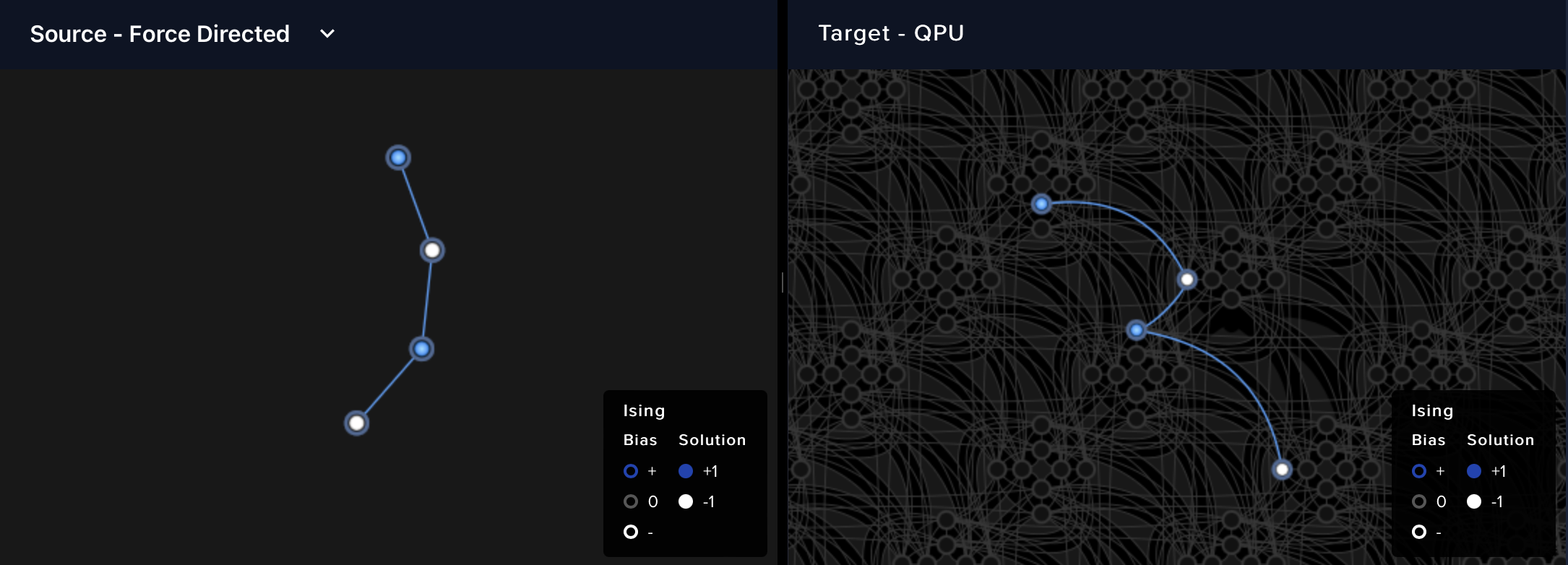}
   \caption{Linear-entangled Ising Model}
   \label{subfig:linear}
 \end{subfigure}

 \begin{subfigure}[b]{0.9\linewidth}
   \includegraphics[width=\linewidth]{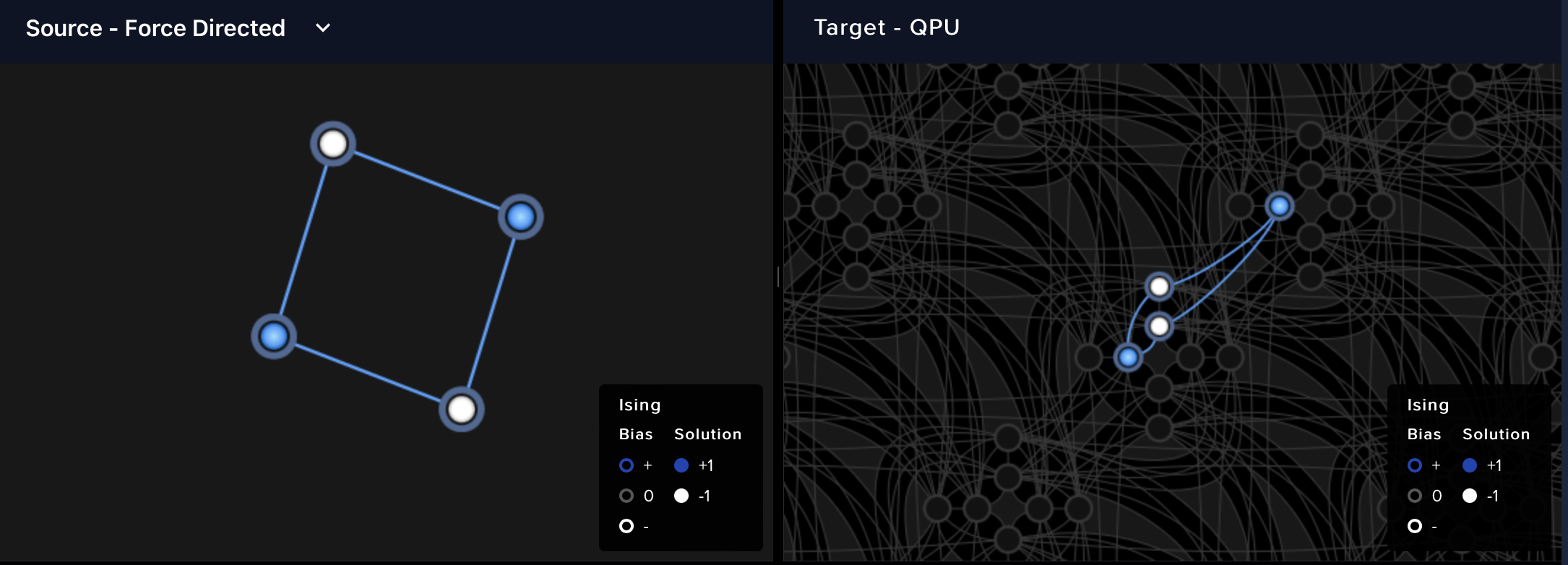}
   \caption{Circular-entangled Ising Model}
   \label{subfig:circular}
 \end{subfigure}

 \caption{Schematics of different Ising Models on a Quantum Annealer, constructed Leap SDK's default inspector tool, depicted across three entanglement configurations. Each subfigure shows the connections among qubits on the left and the implementation on D-Wave's QPU within Pegasus topology on the right.}
 \label{fig:ising_models}
\end{figure}

The implementation of each Ising model variants on the D-Wave system has a sequence of critical steps. The process begins with the initialization phase, where the Ising model parameters, specifically the \(h\) and \(J\) values, are configured to define the problem space accurately. Then, the \texttt{EmbeddingComposite} and \texttt{DWaveSampler} are employed to sample the Ising model multiple times. This step is essential for exploring the solution space thoroughly to identify low-energy states that approximate the ground state of the system. Furthermore, the following graphs for each Ising model variants— full-entangled (Fig.\ref{subfig:circular}), Linear- entangled (Fig.~\ref{subfig:linear}), and circular-entangled (Fig.~\ref{subfig:full})—provide a visual representation of the qubit interactions and their implementation on D-Wave's QPU within the Pegasus topology. Finally, the culmination of this process involves the determination of the ground state by extracting the sample with the lowest energy from the result set, along with its corresponding energy value.

The ground state energy and corresponding qubit configurations for the different Ising models are as follows:

\begin{itemize}
 \item Full-Entangled Model: Achieved a ground state energy of \(-2.0\) with the qubit configuration \(\ket{0, 1, 0, 1}\).
 \item Linear Model: Achieved a ground state energy of \(-3.0\) with the same qubit configuration \(\ket{0, 1, 0, 1}\).
 \item Circular Model: Achieved a ground state energy of \(-4.0\) with the qubit configuration \(\ket{0, 1, 0, 1}\).
\end{itemize}

These results highlight the D-Wave system's capability to efficiently solve the Ising model, with variations in the entanglement patterns affecting the ground state energy and qubit configurations. The ability to visualize the problem setup and solutions via the D-Wave Inspector helps in understanding the quantum annealing process and the dynamics of the Ising model on a quantum annealer.

\subsection{Abstract Mapping For Platform Independent Model}

The primary objective of this study is to establish a conceptual framework and PIM that enables the integration of quantum annealing and gate-based quantum devices for solving the ground state energy for multi-partite quantum systems. PIM standardizes quantum computations for ground state calculation across different platforms like D-Wave's Leap SDK and IBM's Qiskit.

The mapping approach and the implementation of the PIM facilitate a seamless transition between computational paradigms and standardize methodology. Table~\ref{tab:updated_comparative_framework} provides a simplified comparison between the quantum annealing approach and the VQE method in gate-based QC for solving the Ising model:

\begin{table*}[htbp]
\small
\caption{ Comparison of Quantum Annealing and Quantum Gate-Based VQE Approaches in Ground State Calculation}
\begin{center}
\begin{tabular}{|c|c|c|}
\hline
\textbf{Aspect} & \textbf{Quantum Annealer Approach} & \textbf{VQE Approach} \\
\hline
Hamiltonian Mapping & Directly onto hardware qubits & As quantum circuit operators \\
\hline
State Preparation & Physical qubits initialized in superposition & Dynamic state preparation \\
\hline
Entanglement & Fixed by hardware design & Configurable via software \\
\hline
Parameter Optimization & Annealing controls transitions & Algorithmic parameters \\
\hline
Ground State Search & Natural evolution & Iterative optimization \\
\hline
Result Analysis & Immediate hardware output & Computed post-process \\
\hline
\end{tabular}
\label{tab:updated_comparative_framework}
\end{center}
\end{table*}

This comparative table illustrates the distinctive methodologies used in quantum annealing and the VQE method for the specific case, The Ground State Calculation problem. It highlights the shift from a more theoretical discussion to a practical application within the PIM framework, which has been fully realized and now supports these two QC approaches effectively for ground state problem.

\subsection{Platform Independent Model (PIM) for Ground State Calculations}

This paper proposes a PIM for Ground state calculations to achieve standardized representation of quantum computations across various platforms, such as D-Wave's Leap SDK and IBM's Qiskit. The PIM allows users to define quantum systems using high-level parameters, thus eliminating the need for intricate details of the underlying hardware and SDK.

Unlike universal languages, quantum computers require specific instructions tailored to their unique capabilities. These Platform Specific Models (PSMs) act like custom blueprints, accounting for a machine's strengths, like qubit connections, and limitations, such as available operations. While PSMs unlock a quantum computer's full potential, they often demand in-depth knowledge of the underlying hardware, creating a hurdle for programmers.

By contrasting these platform-specific approaches, the need for a PIM becomes evident, as it abstracts these details into a unified model that can be understood and utilized without deep platform-specific knowledge.

\subsubsection{Key PIM Parameters}

The PIM simplifies the quantum system definition by standardizing the following key parameters:
\begin{itemize}
    \item \textbf{Number of Qubits:} Defines the quantum system's size.
    \item \textbf{External Magnetic Field Constant:} Represents the average influence of external magnetic fields across the system.
    \item \textbf{Coupler Constant:} Captures the strength of interactions (couplings) between qubits.
    \item \textbf{Platform Name:} Identifies the targeted QC platform (e.g., 'dwave', 'qiskit').
    \item \textbf{Entanglement Type:} Describes the interconnection pattern of qubits (e.g., linear, circular, full).
\end{itemize}

\subsubsection{Translation Functions between PSM and PIM}

The system implements two primary function types:
\begin{enumerate}
    
    \item[A.] \textbf{PSM to PIM Conversion Functions:}
    These functions (e.g., \texttt{PSM\_to\_PMI\_dwave} and \texttt{PSM\_to\_PMI\_qiskit}) translate PSMs into the platform-independent PIM format. This conversion is crucial for enabling uniform analysis and understanding of quantum systems across different platforms.
    
    \item[B.] \textbf{PIM to PSM Conversion Function:}
    The \texttt{PIM\_ground\_state\_calculator} function takes the platform-independent specifications and generates the necessary platform-specific model. This allows direct execution of quantum simulations or calculations on the designated hardware or simulation framework.
\end{enumerate}

\subsubsection{Use Cases and Benefits of PIM}

The proposed PIM system offers several practical advantages:
\begin{itemize}

    \item \textbf{Platform-Agnostic Quantum System Design:} Users can specify quantum systems using PIM parameters without being limited to a specific platform. This is ideal for theoretical and educational purposes where understanding core concepts outweighs platform-specific details.
    
    \item \textbf{Cross-Platform Translation:} Users working on one platform (e.g., D-Wave's Leap SDK) can effortlessly translate their models to another platform (e.g., IBM's Qiskit) using the conversion functions. This is particularly beneficial for comparing results across platforms or leveraging specific features offered by different platforms.
    
    \item \textbf{Simplified Interface for Quantum Developers:} By abstracting the complexities of each platform into a common set of parameters, PIM simplifies quantum software development. This can accelerate development cycles and reduce the learning curve for new quantum programmers.
\end{itemize}

\pagebreak

\section{Conclusion and Future Work}
\label{sec:conclusion}
\subsection{Key Findings and Contributions}

This study not only proposes a unified framework for integrating different quantum computing (QC) platforms but also successfully implements it for a specific use case of ground state calculation. A significant milestone in this framework is the implementation of the PIM for ground state calculation, which bridges the gap between quantum annealing and gate-based QC. It enables the use of a single set of high-level parameters to define and manipulate quantum systems across different platforms. Additionally, it allows transitions from one PSM to another between the Quantum Annealer model and the Gate-Based Quantum system.

\subsection{Future Research and Potential Impact}

The results outline  possibilities for further research, particularly in refining these mapping techniques for even more sophisticated and broader QC challenges. Future work could explore extending the PIM to support emerging quantum technologies and integrating Domain-Specific Languages (DSLs) that could further simplify quantum algorithm development across platforms.

\subsubsection{Model Enhancements }

To further improve the PIM system, several aspects can be considered:

\begin{itemize}
\item \textbf{Validation and Error Handling:} The conversion functions should be robust in handling exceptional cases and provide informative error messages for unsupported configurations or invalid inputs.
\item \textbf{Support for Additional Platforms:} As the field of QC evolves, incorporating support for emerging platforms will ensure the continued relevance and utility of the PIM system.
\item \textbf{User-Friendly Documentation and Examples:} Comprehensive documentation and illustrative use cases will empower users to use the PIM system effectively. Tutorials or interactive web applications could significantly enhance user engagement and understanding. 
\item \textbf{Extending Compatibility:} Currently, the framework is tailored only for Ising model solutions for the ground state problem. To fully tap into the potential of quantum technologies, it is essential to expand its applicability to encompass all programming algorithms in quantum computing. 
\end{itemize}

These improvements will solidify MDE frameworks as  valuable tools for facilitating standardized development and execution of quantum computations across diverse platforms for quantum or hybrid applications.

The impact of using MDE principles in QC is profound, potentially impact how quantum algorithms are developed and making QC more accessible to a wider audience.

\newpage

\section*{References}

\bibliographystyle{IEEEtran}
\bibliography{references}
\end{document}